\documentclass{aastex6}
\usepackage{amsmath}
\usepackage{graphicx}
\usepackage{epstopdf}
\usepackage{natbib}
\usepackage{hyperref}

\shorttitle{Separated Fringe Packet Binaries III}
\shortauthors{Farrington et al.}

\begin{document}

\title{Separated Fringe Packet Observations with the CHARA Array III. \\
The Very High Eccentricity Binary HR~7345}


\author{C.D. Farrington}
\affil{The CHARA Array, Mt. Wilson Observatory, Mt. Wilson, CA 91023}
\email{farrington@chara-array.org}
\author{Francis C. Fekel}
\affil{Tennessee State University, 3500 John A. Merritt Blvd, Box 9501, 
Nashville, TN 37209}
\author{G.H. Schaefer, T.A. ten Brummelaar}
\affil{The CHARA Array, Mt. Wilson Observatory, Mt. Wilson, CA 91023}

\begin{abstract}
After an eleven year observing campaign, we present the combined 
visual--spectroscopic orbit of the formerly unremarkable bright star 
HR~7345 (HD~181655, HIP~94981, GJ~754.2). Using the Separated Fringe Packet (SFP) method with the CHARA Array, we were able to determine a difficult to complete orbital period of $331.609 \pm 0.004$ days. The 11 month period causes the system to be hidden from interferometric view behind the Sun for 3 years at a time.  Due to the high eccentricity orbit of about 90\% of a year, after 2018 January the periastron phase will not be observable again until late 2021. Hindered by its extremely high eccentricity of 0.9322 $\pm$ 0.0001, the double-lined spectroscopic phase of HR~7345 is observable for 15 days.
Such a high eccentricity for HR~7345 places it among the most eccentric systems in catalogs of both visual and spectroscopic orbits. For this system we determine nearly identical component masses of 0.941 $\pm$ 0.076 $M_{\odot}$ and 0.926 $\pm$ 0.075 $M_{\odot}$ as well as an orbital parallax of 41.08 $\pm$ 0.77 mas. 
 
\end{abstract}


\vskip 0.5pt

\section{Introduction}

\subsection{Observational History of HR~7345}
HR~7345 (HD~181655, HIP~94981, GJ~754.2) is a 6th magnitude star in the 
constellation Lyra near the boundary line with Cygnus and seemed for
decades to be an unremarkable system aside from it's proximity to the Sun. We discovered it to be a highly eccentric binary, even though it had not been identified in previous multiplicity surveys.
It was included as part of a David Dunlap Observatory (DDO) radial 
velocity survey of 681 relatively bright stars for which velocities were 
lacking \citep{young45}. A decade later \citet{HALI55} used the same 
DDO spectra to calculate the luminosity and spectroscopic parallax 
($\pi_{sp}$ = 79.7 mas) for the system and classified its spectrum as 
G8~V. Two years later, \citet{CRIS57} determined its trigonometric 
parallax ($\pi_{trig}$ = 39 mas) from photographic plates and found a 
value very close to the modern Hipparcos \citep{Gaia16} measurement ($\pi_{Hip}$ = 39.34 mas). 
During the next 20 years, it was spectroscopically and photometrically 
measured and classified, with no hint of variability, even being 
listed as a radial velocity standard star by \citet{BE79} from 20 
measurements over three years at the Fick Observatory.  The system was 
measured six more times between 1978 and 1983 at the McDonald 
Observatory 2.1-m telescope where it just barely fell outside of the 
2.5~$\sigma$ error limit for their definition of a ''radial-velocity 
standard star" \citep{BA86}. HR~7345 was even observed in the early 
eighties by the primary author's thesis advisor with speckle interferometry \citep{McA1987} on the Kitt Peak 4~m telescope, which gave a null result for their single observation in 1985. The timing of that 
measurement was particularly unlucky, as the companion would have been 
just coming out of periastron but not yet separated enough for easy 
resolution on the 4~m telescope. Radial-velocity data were again 
acquired during the CORAVEL survey of \citet{DM1}, who searched for 
companions of solar-type stars in the solar neighborhood. Twelve 
measurements of HR~7345, taken over the course of 1200 days between 
1983 and 1989, indicated very little velocity variation.  In retrospect, 
their timing was most unfortunate, as they were tantalizingly close to 
the very short observing window when the system would have exhibited 
double lines between 1983 to 1985. Unfortunately, again because of the 
very limited window, nearly all subsequent observations of HR~7345 
failed to show evidence of binarity \citep{Duf95, Feh97, Nid02, Gray03, 
Halb03, Nord, Val05, Holm09, Crif10, Soub13, Gaia16}.  In fact, out of 
all the radial velocity observations collected during the past 50 years, 
only one, an ELODIE spectrum acquired in 2000 November, was close 
enough to periastron to exhibit partial separation of its double lines 
\citep{Prug07}. Finally, the Palomar-Testbed Interferometer observed the system over 40 times 
between 1998 and 2005 and saw no evidence of a companion with their 
86-110m baselines and deemed it to be a suitable calibrator star 
\citep{GVB07}. Further inquiry into the reasons why it was not detected are ongoing, but due to orbital elements projected backwards, there were several years when the companion should have been detectable. 

\subsection{High Eccentricity Binaries}
As is often the case, many of the most interesting systems are discovered by accident. Although not originally considered until many observations were obtained, the importance of surveying high-eccentricity binary systems cannot be overstated. Characterizing high-eccentricity systems can provide insight into the statistics of stellar formation mechanisms, multiplicity fractions and star formation rates which all lead to the Inital Mass Function (IMF) \citep{AMB1937, BATE2009, TOKO2016b}.  It is well known that both visual and spectroscopic observations can easily miss a significant fraction of high-eccentricity systems \citep{DM2, RAG2010, GRI2012} by something as simple as timing where either the relative motion of the components in the visual case is very slow for most of their orbital period, or the relatively short time when the spectra would exhibit double-lines.  This combined with the wide variation of inclinations and distances means even discovering high-eccentricity systems is often left up to a chance observation. Once found, these systems can help define the limits of the eccentricity distributions \citep{RAG2010}, relations between period and eccentricity \citep{FIN1936}, and the mechanism that creates such extreme systems (See Section \ref{HEBC}).

\section{Observations}
During the primary author's dissertation research, one of the first 
systems observed was HR~7345 in a search for companions that were missed 
in the multiplicity survey of \citet{DM2}. It was suspected that, prior 
to the Hipparcos mission, there could well be systems in the afore 
referenced multiplicity survey that were not within the $22$ pc distance
 parameter as well as the possibility of undiscovered companions that 
were below the resolution capability of single aperture interferometry. In this case, HR~7345 was both the first 
system in that survey to exhibit multiple fringe packets and the first 
previously unknown companion discovery. Unbeknown to us, its very high 
eccentricity and nearly one year orbital period closely aligned periastron 
with its conjunction with the Sun from 2005-2010.  Thus, after five years 
of observations, all the measurements were loosely clustered in the N-S 
direction with very little movement in position angle or separation.  
The following year, we were able to catch a fast moving phase in the 
E-W direction that returned our interest to the system, and HR~7345 was 
added to a list of systems to be monitored more often.  As its components 
are of nearly equal brightness in the infrared, the fringe packet 
amplitudes are nearly identical, and so, it was not until the end of 2015 
that we identified the correct period with the help of spectroscopic 
observations. By the time we were able to pinpoint the time of 
periastron passage, we were unable to acquire interferometric observations 
during that fast moving phase due to unfavorable weather conditions at 
Mount Wilson during 2016 March and 2017 February. Fortunately, during predicted periastron passage on January 24, 2018, we were able to use the three longest baselines of the CHARA Array in the last hour of the night when the star was just $16\deg$ above the horizon to get three observations to pin down the unobserved part of the orbit.  

\subsection{Interferometric Observations}
We obtained observations of HR~7345 from 2005 October through 2018 January 
with multiple combinations of baselines using the Classic and CLIMB 
\citep{CLIMB2} beam combiners of the CHARA Array at Mount Wilson Observatory 
in southern California \citep{CHARA2}. During the first three years, the Separated Fringe Packet (SFP) project (\citet{FAR2010} and \citet{FAR2014}, hereafter referred to as Paper I and Paper II, respectively)
only had access to the Classic beam combiner, which 
could only observe with one baseline at a time and took significant time 
to reconfigure.  As such, there were sometimes gaps of one to several 
days between different baseline measurements.  Luckily, during this period 
the observation windows lined up with the very slow moving apastron phase. 
With the advent of CLIMB in 2009, we were able to collect data on three 
baselines within fifteen minutes of acquisition. Data were routinely taken 
with the CHARA Array's outer telescope triangle (S1-W1-E1), as they have the 
greatest separation and are able to probe the smallest separations. The setup 
of the Array in general, the SFP method, error sources, and the 
acquisition/reduction of data are discussed in detail in Paper I.  The 
conversion of reduced data into ''on-sky 2D" measurements is expanded upon 
and described in Paper II. For the final astrometric measurement, the predicted separation was far below the resolution limit of the SFP process, so three calibrated brackets were taken using the traditional visibility method with the same triangle mentioned previously and using HD~174602 and HD~173649 as calibrators with a C1-O-C2-C1-O-.. sequence and calibrator diameters of 0.356 and 0.388~mas respectively \citep{JMMC}.  Data were reduced using a pipeline developed by J.~D. Monnier, using the general method described in \citet{MONNIER2011} and extended to three beams (e.g., \citet{KLUSKA2018}). The calibrated OIFITS file is available through the Optical Interferometry DataBase (OIDB)\footnote{OIDB searchable database located at http://oidb.jmmc.fr/index.html} or upon request. 

All 1-D observations from the CHARA Array and subsequent 2-D calculations from 2005-2018 are listed in Table \ref{tab_sfpobs}. The first four columns characterize the 1-D measurements taken by a single baseline (Time, baseline length, fringe separation, and position angle of the baseline), while the last six columns are the average position of the detected companion with associated errors. The conversion of time frames in the averages is to consolidate to one reference frame congruent with the spectroscopic observations described in the following section. The final line of Table \ref{tab_sfpobs} contains the periastron observation taken during 2018 January. We solved for the binary position on UT 2018 Jan 24 using the adaptive grid search procedure described in \citet{schaefer16}.  We derived a flux ratio of 1.07 $\pm$ 0.02 in the K-band at the position listed in the last line of Table \ref{tab_sfpobs} 
   
\subsection{Spectroscopic Observations}
We obtained observations of HR~7345 from 2014 June through 2018 January 
with the Tennessee State University (TSU) 2~m automatic spectroscopic 
telescope (AST) and a fiber-fed echelle spectrograph \citep{ew07}. That 
telescope is situated at Fairborn Observatory near Washington Camp in 
southeastern Arizona. The detector is a Fairchild 486 CCD having a 
4096$\times$4096 array of 15~$\mu$m pixels. For our observations we 
used a 200 $\mu$m fiber that results in a spectral resolution of 0.24~\AA. 
The signal-to-noise ratio of the spectra was typically about 90per resolution elementat 6000~\AA. \footnote{The echelle spectra	FITS format files are available at http://ast2.tsuniv.edu/HD\_181655/}

\citet{ftw09} provided an extensive general description of velocity 
measurement of the Fairborn AST spectra. In the case of HR~7345 we used 
a solar line list that covered the wavelength range 4920~--7100~\AA\ 
and fitted the individual lines with a rotational broadening 
function \citep{lf11, fg11}.  Because the orbit of HR~7345 has an 
extremely high eccentricity, the lines of its two nearly identical 
components appear as highly blended, single features for the vast 
majority of its orbit, and so a single velocity was determined for 
most of those observations. The maximum velocity separation in that part of the orbit
occurs at about phase 0.75 and is just 4 km~s$^{-1}$. However, once the velocities of the two 
components began to change significantly near periastron passage, 
causing the single lines to broaden and weaken in strength, we 
determined velocities for both components by fitting the still very 
blended profiles with two rotational broadening functions. The 
starting values for the depths and widths of the components in the 
blends were determined from those values found for well separated 
lines at phases very close to periastron. Velocities from the AST CCD 
spectra have a zero-point offset of $-$0.6 km~s$^{-1}$ relative to the absolute radial velocities cataloged in \citet{s10}.  Thus, we added  0.6 km~s$^{-1}$ to our measured 
velocities.  All of the spectroscopic observations for this system are 
listed in Table \ref{tab_SBobs}. 
  
\section{Orbital Solution}

\subsection{Interferometric Orbit}
\label{sect.vb}

We determined an initial set of orbital parameters by fitting the 44 
interferometric positions using the three-dimensional grid search 
procedure described by \citet{schaefer06}.  We then refined the orbital 
parameters with a Newton-Raphson method to minimize $\chi^2$ between 
the measurements and the orbit fit by calculating a first-order Taylor 
expansion for the equations of orbital motion.  We adopted an iterative 
approach to adjust the weights of each measurement.  First, we uniformly 
scaled the uncertainties on the positions to force the reduced 
$\chi^2_\nu$ to equal unity (where $\nu$ is the degree of freedom).  If 
the residual from any measurement compared with the orbital fit was more 
than three times the measurement error, we adjusted the weight of that measurement such that the uncertainty of the data point increased by a factor of 10.  We then re-computed the orbital fit, 
uniformly re-scaled all of the uncertainties to force the reduced 
$\chi^2_\nu$ to equal unity, and compared the residuals to adjust the 
individual weights again.  We repeated this process until no more 
uncertainties were adjusted.  In the end, a total of five measurements 
had their uncertainties adjusted. The final scaled uncertainties adopted for all measurements are reportedin columns eight and ten of Table~\ref{tab_sfpobs}.

\subsection{Spectroscopic Orbit}
Our very first AST spectrum showed partially separated double lines that 
had nearly equal depths, but without sufficient knowledge of the orbital 
elements, our next spectrum was not obtained until nearly two weeks 
later, by which time the partially blended, double-lined profile had 
become a narrow single-lined profile. A spectrum 342 days later showed 
a broadening and weakening of the single-lined profile but that was the 
last spectrum obtained before monsoon season. Nearly 330 days later 
there was again an indication that modest velocity changes had occurred, 
when one spectrum showed the component lines partially resolved. Despite the 
very limited number of spectra with indications of velocity changes, the 
very extensive number of single-lined spectra suggested a high 
eccentricity orbit with a period of about 330 days rather than a much 
more circular orbit with twice that period.

With the results of a preliminary joint astrometric--spectroscopic 
orbital solution, preparations were made to attempt spectroscopic 
observations during the next periastron passage that was to occur in the
latter half of 2017 February. However, given the near 11 month period, 
that predicted periastron passage would occur less than 50 days after 
HR~7345 reached the same right ascension as the Sun. In most cases 
observations so close to the Sun are precluded.  However, when the Sun 
and HR~7345 have the same right ascension, they have a declination 
difference of 59\arcdeg. This large difference enabled us to acquire 
spectra of HR~7345 during the periastron passages 
of 2017 February and 2018 January.

Despite the short observing window at the very end of the nights in 
2017 February, we attempted to get multiple observations of 
HR~7345 each night for the two week period around predicted periastron. 
Although the night of maximum velocity separation was cloudy, we were 
able to obtain spectra on the adjacent nights. However, having 
missed the night of maximum velocity separation in 2017 February,
we decided to obtain spectra of the system at its next periastron passage
in 2018 January when it was even closer to the Sun. Fortunately, the
weather cooperated, and we successfully acquired multiple spectra of
HR~7345 on nearly every night when the double lines were resolved 
including the night of maximum velocity separation. Thus, the spectroscopic 
solution is well constrained.

We first determined spectroscopic orbits of the individual components 
with SB1, a program that iterates the orbital elements by differential 
corrections \citep{betal67}. To obtain a simulaneous two-component 
solution, we used SB2, which is a slightly modified version of SB1.

\begin{figure}
	\begin{center}
		\scalebox{0.5}{\includegraphics{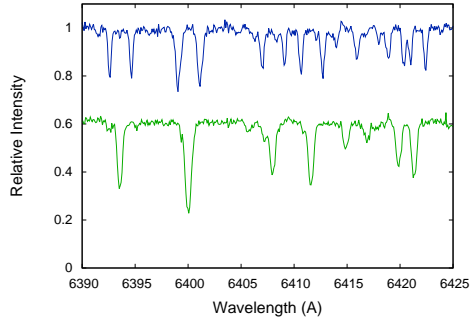}}
		\caption{A portion of an echelle order for
			HJD 2458142 (upper, double-lined spectrum) and HJD 2458153
			(lower, single-lined spectrum). The upper spectrum is our
			observation closest to maximum velocity separation. The lower
			spectrum, shifted downward for clarity, has an orbital phase of
			about 0.03 and has a velocity separation of 13 km~s$^{-1}$.
			The latter spectrum is representative of spectra with very
			blended lines from which we obtained useful velocity
			measurements. Spectra with the components having smaller
			velocity separations had systematic residuals and were not
			used in our orbital solutions.}
	\end{center}
	\label{SBObs}
\end{figure}

\begin{figure}
	\begin{center}
		\scalebox{0.5}{\includegraphics{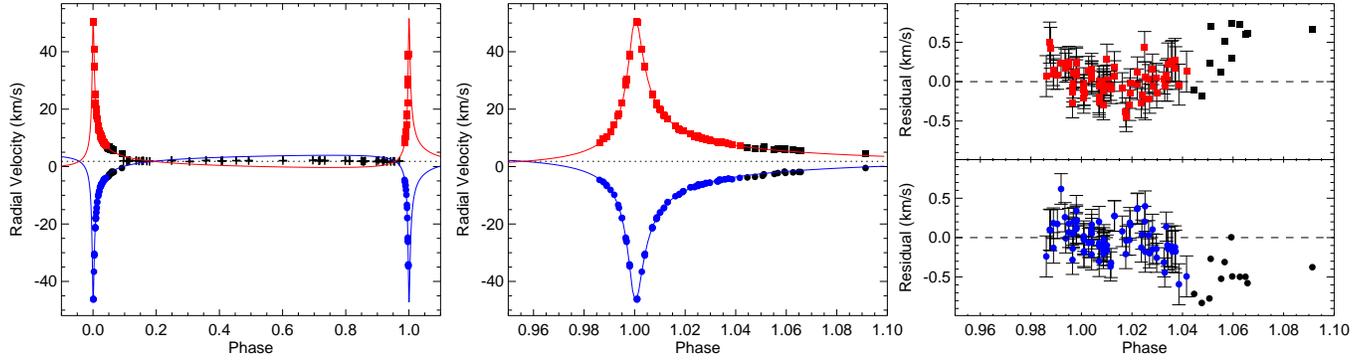}}
		\caption{Radial velocity measurements for HR 7345 A (blue circles) and B (red squares). The solid line is the best fitting orbit computed from a simultaneous fit to the CHARA measurements and the spectroscopic radial velocities. The center-of-mass velocity is shown as a dotted line.  The filled black symbols are the individual velocities for the two components that were given zero weight because the lines of the two components were severely blended.  The black crosses show the single-lined velocity measurements.  The middle panel shows a zoomed in view of the phase-wrapped radial velocity peak.  The panels on the right show the residuals between the measured radial velocities and the orbital fit for each component.}
	\end{center}
	\label{SB}
\end{figure}

Because of its very high eccentricity, the orbital fit is dominated by 
the velocities near periastron that are determined from component lines 
that are at least partly resolved at our resolution. This is a window 
of about 15 days and so covers only about 5\% of the orbit. We tried 
various solutions with different weights for those velocities that 
were determined for the completely blended double lines. In the final 
spectroscopic solution we assigned zero weight to the velocities of 
those observations that were measured as single-lined. However, for 
the spectra having broadened but completely blended lines, the velocity 
weighting was more problematic. Velocities for many of the spectra with very blended lines that were measured
as double lined show systematic velocity residuals (Fig.~\ref{SB})
and so were not used in the final solution. In the final orbit all velocity 
measurements, whether single or double, between phases 0.042 and 0.985 
were given zero weight. Radial velocities from the 12 most blended remaining
observations (see Figure \ref{SBObs}) were given weights of 0.5, while all other velocities
were given unit weights. The solution given by only the spectroscopic data is listed in column two in Table \ref{orbittab}.

\subsection{Combined Solution}

\begin{figure}
	\begin{center}
		\scalebox{0.64}{\includegraphics{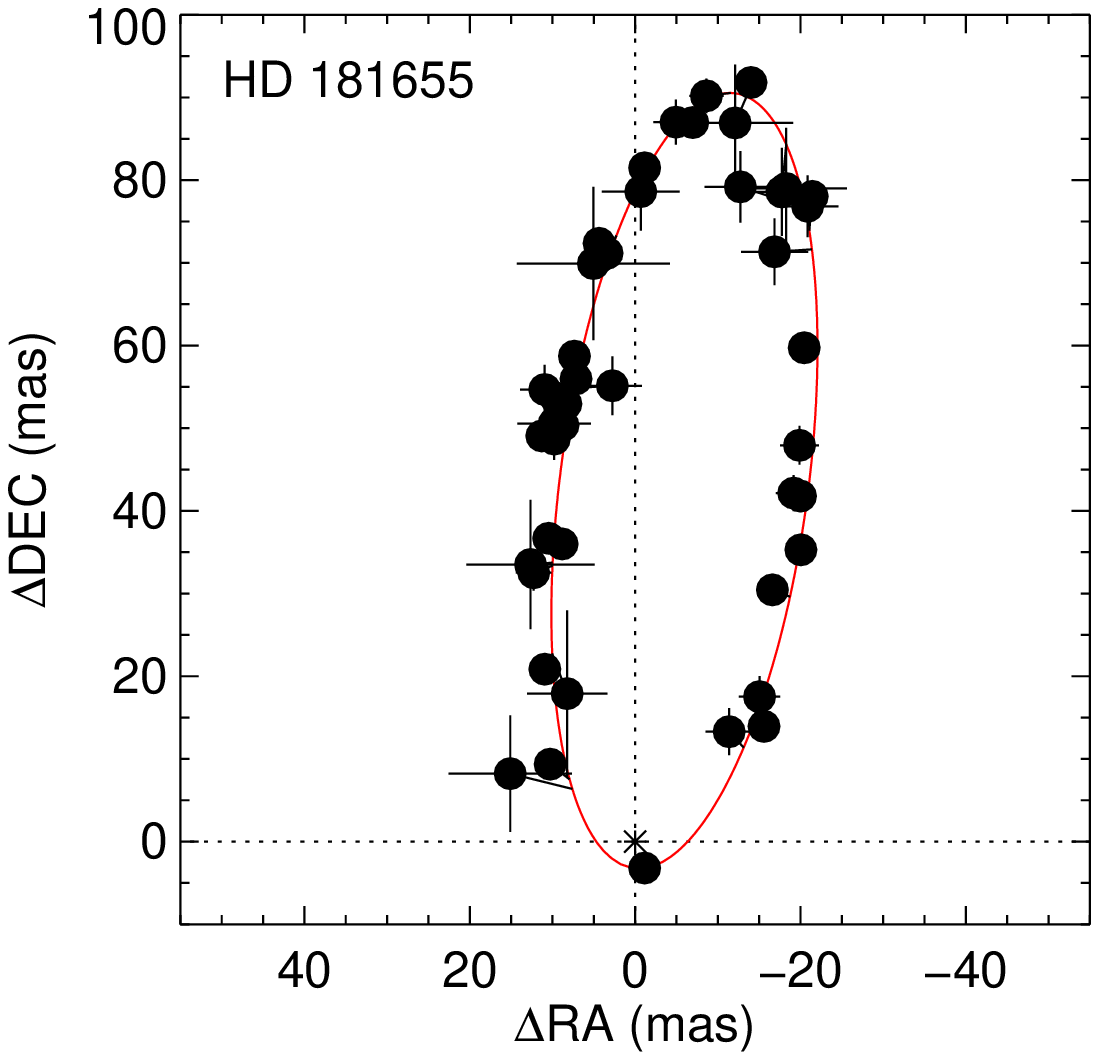}}
		\scalebox{0.48}{\includegraphics{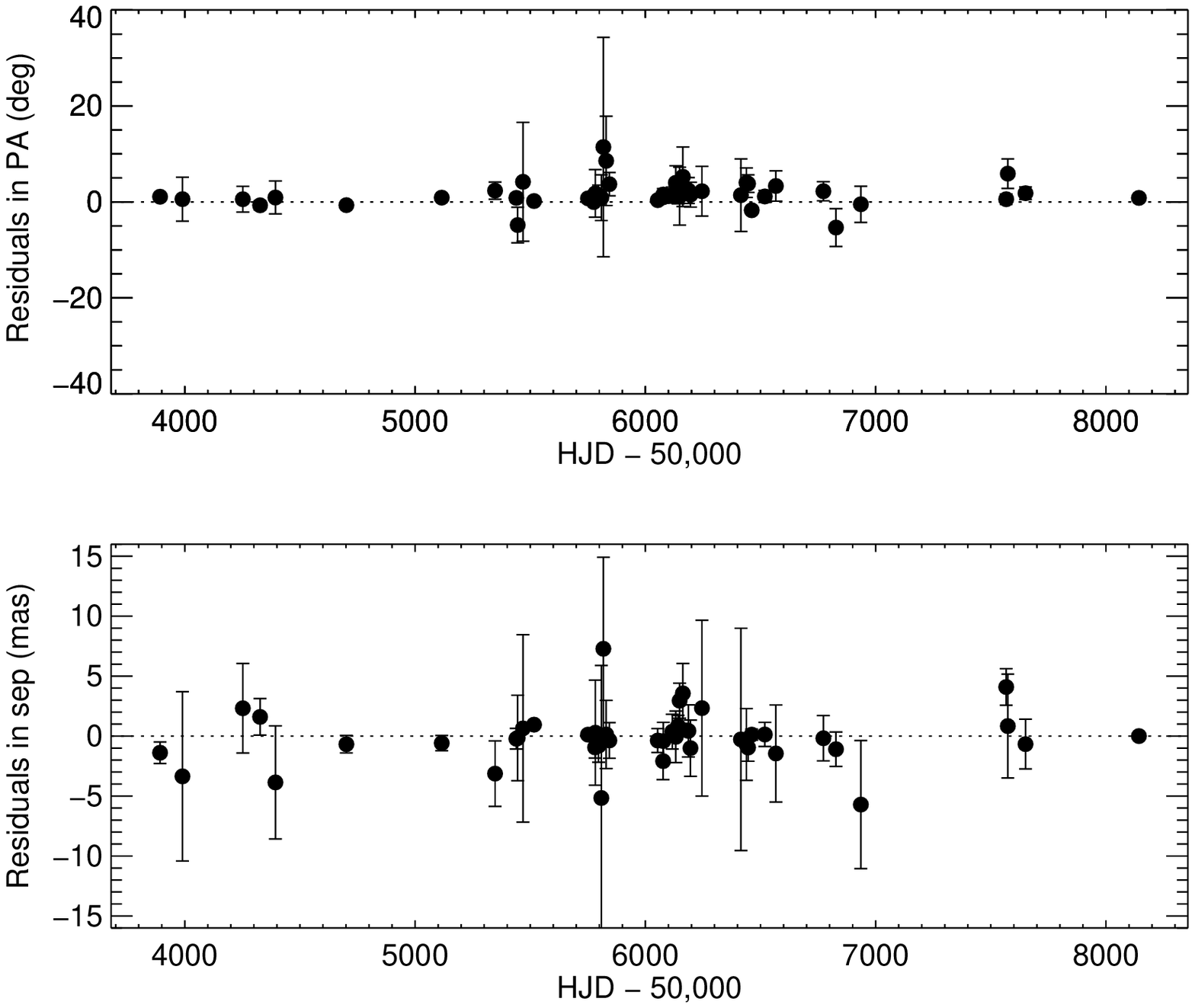}}
		\caption{Left: Orbital motion of HR~7345 A relative to B as 
measured with CHARA Classic and CLIMB.  Overplotted in red is the best fit 
orbit computed from a simultaneous fit to the CHARA measurements and the 
spectroscopic radial velocities.  Right: Residuals between the measured 
position angle and separation compared with the best fitting orbit.}
	\end{center}
	\label{VB}
\end{figure}

We also fit an orbit simultaneously to the visual and spectroscopic data.  
In the joint fit, we applied the measurement weights determined based 
on the individual orbit fits.  To give equal weight to each set of data, 
we scaled the uncertainties from each set to force the reduced  
$\chi^2_\nu$ to equal unity.  We then computed the simultaneous orbital 
fit following a similar Newton-Raphson technique as described in 
Section~\ref{sect.vb}, but expanded to fit all 10 orbital parameters.  
The parameters from the joint fit are listed in the last column 
of Table~\ref{orbittab}.  The uncertainties were computed from the 
diagonal elements of the covariance matrix.  Figures \ref{VB} and \ref{SB} 
show the measurements compared with the joint orbital fit.

\section{Orbital Parameters}

The parameters from the joint orbital fit are mostly consistent within 
their uncertainties with those determined from the individual fits to 
the visual and spectroscopic data.  The values for $P$, $T$, $e$, and 
$\omega_A$ are constrained more strongly by the radial velocity 
measurements. The changes in $i$, $\Omega$, and $\omega_A$ in the visual 
only fit compared with the joint fit are a result of the tighter 
constraints on $e$ and $\omega_A$ provided by the radial velocities.
Table~\ref{orbparam} gives the masses and distance computed from the 
orbital parameters.  Our combined orbital solution after the 
2017 February \textbf{periastron} passage produced an inclination of 32.4$^\circ$ 
$\pm$ 1.3$^\circ$ and individual stellar masses of 0.73 and 0.72 $M_{\odot}$. 
Such values are significantly lower than the value of $\sim$0.95 $M_{\odot}$ 
expected for a spectral type of G5~V \citep{Gray03}. However, the dynamical 
masses are highly sensitive to the low inclination.  
After obtaining interferometric observations as the system passed 
through the 2018 periastron, the revised orbit produces 
masses of $0.941~\pm~0.076$ and $0.926~\pm~0.075M_{\odot}$, which are very close to the value 
expected for a G5~V star. The orbital parallax of 41.1~$\pm$~0.8 mas is larger than the Hipparcos 
parallax of 39.39~$\pm$~0.33 mas, which has no flag for binarity listed 
in any of the revisions of the data set \citep{HIP2}. Similarly, the Gaia DR2  \citep{Gaia16} parallax (39.646~$\pm$~0.098~mas) is closer to the Hipparcos values rather than the orbital parallax.  While there is not a direct flag in the Gaia DR2 \citep{LURI2018}, the data collected does show abnormally high astrometric $\chi^2$, ``Goodness of Fit", and excess noise values indicate a statistically poor fit, likely due to the not-quite equal magnitude of the system's components.  

\section{High Eccentricity Binaries Comparisons} \label{HEBC}

Comparing the orbital elements for HR~7345 to systems listed in the 
Ninth Catalogue of Spectroscopic Binary Orbits \citep{9TH} and the 
Sixth Catalog of Orbits of Visual Binary Stars \citep{6TH}, one 
immediately sees that HR~7345 stands out. Sorting by eccentricity, it 
ranks as the 12th most eccentric in the spectroscopic catalog, and 
is 83rd in the visual binary catalog, but of those systems that are 
more eccentric, none has such a short period.  The closest comparable 
systems in the visual catalog have periods on the order of 800 days 
(HD~66751 and HD~212029). In the spectroscopic catalog the shortest 
period system with a larger eccentricity than HR~7345 is HD~137763
which has a period of 890 days, while with a period of 298 days, 
HD~89707 is a system with a similar period and very high but smaller 
eccentricity than HR~7345.  Both of the visual systems have astrometric 
solutions from Hipparcos data, but the binaries were published with 
two differing orbits for each system with discrepancies between the 
eccentricities (e = 0.97$^{+0.02}_{-0.56}$ and e = 0.56$^{+0.43}_{-0.27}$ 
for HD~66751, e = 0.92$^{+0.07}_{-0.34}$ and e = 0.99$^{+0.00}_{-0.19}$ 
for HD~212029) in two different tables in the same paper \citep{GOLD06}. 
For the spectroscopic systems, HD~89707 is only single-lined and its
companion is a brown dwarf candidate. The initial orbit from \citet{DM2}
has an eccentricity of 0.927$\pm$0.014 but a more recent orbit by 
\citet{SAH} reduced it to 0.900$^{+0.039}_{-0.035}$. However, even this
more recent orbit remains questionable as the rapid nodal passage is
very poorly covered. Likely the best system for comparison is HD~137763, 
which has both a visual and spectroscopic solution with a very similar 
''extreme orbit" paper espousing the superlative eccentricity of 
0.976 \citep{STRA} from high precision spectroscopy followed by a 
visual and combined solution from \citet{Toko2016}. While the 
eccentricity of HD~137763 is more extreme, the period of HR~7345 is 
significantly shorter while still very eccentric.

\citet{GRI2} reviewed the limited number of spectroscopic binaries with 
published eccentricities greater than 0.9 that were known at that time. 
He then extensively discussed the great difficulty of identifying such 
systems spectroscopically pointing out that the discoveries are 
influenced by observational selection effects and the pure luck of 
observing a system in such a small phase window of the orbit.  As 
discussed in Section 1, of the dozens of spectroscopic observations 
acquired before our spectroscopic observations began, only one was at 
a phase that showed partially resolved components. Thus, in the case 
of our efforts on HR~7345, its binary nature was first detected 
interferometrically rather than spectroscopically. 

\citet{GRI2} goes on to mention that there is a second selection effect, 
as the longitude of periastron of previously identified high eccentricity 
binaries is clustered around 90\arcdeg and 270\arcdeg, because the slow 
velocity change resulting from those longitudes of periastron provides 
a much wider observational window to catch a double-lined phase than the 
perpendicular case. The lone ELODIE observation of HR~7345 with its 
partially resolved components \citep{Prug07} produced no follow-up campaign 
to establish its orbit. Thus, it is indeed fortunate that the system was 
independently discovered to be a binary by visual means. Even knowing 
that HR~7345 is a binary, it was quite difficult over the course of 
12 years to determine the correct visual orbit for a nearly equal 
component binary having almost a one year period. Indeed, the 
determination of an accurate periastron only occurred after a decade of 
periodic observation and with the most recent periastron occurrence 
barely observable before the system goes into hiding behind the Sun 
for the next three years.

Given the extreme eccentricity of HR~7345, the original architecture of this binary system was almost certainly very different than it is now.  There are several mechanisms that could produce such large eccentricities in a nearly equal mass binary, all of which involve a third or fourth component.  Without the ability to probe the initial conditions, it is hard to say how many components the system originally had during formation.  However, the most probable explanation for just the eccentricity extreme is likely the Kozai-Lidov mechanism (see \citet{K62}, \citet{L62}), where the system was previously a non-coplanar triple with a less massive wide component, likely smaller than the two other components with sufficient inclination to the inner orbit to cause oscillations in the eccentricity increasing to current levels and then being ejected from the system by a close encounter \citep{ANA1986}. During a recent check of proper motions and distance of the surrounding area after the Gaia DR2 was published earlier this year (HR~7345: $\mu_\alpha$=$-61.524 \pm 0.191$~mas~yr$^{-1}$, $\mu_\delta$=$-183.668~\pm~0.211$~mas~yr$^{-1}$), an amended entry was placed at the top of the list. Convienently, this is a candidate (2MASS~J19193649+3720077) for this interaction at only $35$ arcsec in angular distance. Previously listed as only an X-ray and IR source, the Gaia DR2 added distance, proper motion, and magnitude ($\pi$=$38.88 \pm 0.23$~mas, $\mu_\alpha$=$-53.47 \pm 0.51$~mas~yr$^{-1}$, $\mu_\delta$=$-204.55 \pm 0.49$~mas~yr$^{-1}$, and V-mag=$11.484$ K-mag=$7.87$) very similar to that of HR~7345. Due to it being an X-ray emitter, moderately bright in the IR, and faint in the visible, the object is likely to be chromospherically active M-dwarf which fits the speculated criteria for the instigator of the Kozai-Lidov mechanism. An ejection of this type would certainly explain the rare high eccentricity. Further investigation of the components and the formation mechanisms could be followed up in further studies.

\section{Conclusion and Discussion}
Prior to the observations taken in 2018 January, the calculated 
orbit left many open issues because of the unobserved part of the 
orbit around periastron.  The value and large error in inclination due 
to the missing part of the orbit produced masses that were significantly 
lower than expected, but a large enough error bar to include the 
canonical masses for stars of that spectral classification. 
While the data and orbit were deemed solid enough to publish as they were, 
we decided to make one more attempt if the weather permitted to 
improve the combined orbit.  The weather over Fairborn Observatory
proved to be significantly more cooperative in obtaining spectra of the 
system during the 2018 periastron and filled in the missing single 
day of phase coverage. Observations with the CHARA Array were 
setup for the four days on either side of the predicted periastron passage, 
but due to wind, clouds, and humidity, we were only able to obtain a 
single data set directly on periastron. Fortunately, the results
from this single night proved to be very valuable and allowed a final 
recalculation of the orbit, which provided significant reductions in 
the inclination value and associated errors. While the 2018 
observations significantly improved the orbit and resulting masses, those 
masses, although consistent with the spectral type, have relatively large
uncertainties.

 With the use of our orbital period and component masses, we obtained
 from Kepler's third law a semi-major axis of 1.156 AU. The system's
 large eccentricity produces a periastron separation of 0.079 AU or
 16.9 $R_{\sun}$. \citet{h81} has shown that for stars in an eccentric
 orbit the rotational angular velocity of an individual star will
 tend to synchronize with that of the orbital motion at periastron.
 He called this situation pseudosynchronous rotation. To see whether
 the components of HR~7345 have achieved that state, we first
 determined the projected rotational velocities of the components
 from our rotational broadening fits and found $v$~sin~$i$ values
 of 2.6 $\pm$ 1.0 km~s$^{-1}$ for each component. If the rotational
 and orbital axes are parallel, as is generally assumed for stars
 in binary systems, then our orbital inclination of 29.5\arcdeg
 produces rotational velocities of 5.3 $\pm$ 1.0 km~s$^{-1}$. From
 Kepler's third law, the pseudosynchronous period is 5.88 days.
 If we adopt radii of 0.95~$R_{\sun}$ for the G5 dwarfs, then the
 predicted pseudosynchronous velocities of the components are
 8.2 km~s$^{-1}$. Thus, although the rotational velocities that
 we have determined for the two components of HR~7345 are larger
 than typical values of about 2--3~km~s$^{-1}$ for most late-type
 stars \citep{Val05}, the components of HR~7345 have not
 yet reached pseudosynchronous rotation.

In a review of precise masses and radii for normal stars, 
\citet{Torres10} cataloged 23 systems with both visual and spectroscopic
orbits that produced component stellar masses determined to better
than 3 percent. As listed in Table \ref{orbparam}, the masses of HR~7345 have a 9 
percent uncertainty. As most of the error in the masses is tied to the inclination, more observations around periastron 
should refine the orbit further and decrease the mass uncertainty. With the current orbital elements, the ideal time to observe the system is the during the week before and after periastron, when the system is separated by less than 15~mas. Such
observations will be attempted around 2021 September 11, during the next easily observable periastron passage.

\acknowledgments

This work is based upon observations obtained with the Georgia State 
University Center for High Angular Resolution Astronomy Array at Mount 
Wilson Observatory.  The CHARA Array is supported by the National 
Science Foundation under Grants No. AST-1211929 and AST-1411654.  
Institutional support has been provided from the GSU College of Arts 
and Sciences and the GSU Office of the Vice President for Research and 
Economic Development. 
The research at Tennessee State University was supported in part by
the state of Tennessee through its Centers of Excellence program.
This research has made use of the SIMBAD database, operated at CDS, 
Strasbourg, France. Thanks are also extended to Ken Johnston and the 
U.\ S.\ Naval Observatory for their continued support of the Double 
Star Program and the \citet{WDS}. We would also like to thank John Monnier for the use of his CLIMB data reduction pipeline that made reduction of the periastron data possible, and Brian Mason for the significant work on the preliminary orbits of this system over the past ten years.
This work has made use of data from the European Space Agency (ESA) mission
{\it Gaia} (\url{https://www.cosmos.esa.int/gaia}), processed by the {\it Gaia}
Data Processing and Analysis Consortium (DPAC,
\url{https://www.cosmos.esa.int/web/gaia/dpac/consortium}). Funding for the DPAC
has been provided by national institutions, in particular the institutions
participating in the {\it Gaia} Multilateral Agreement.

\clearpage
\begin{deluxetable}{rrrrrrrrrr}
	\tablewidth{0pt}
	\tabletypesize{\footnotesize}
	\tablecaption{Visual Binary Measurements for SFP System HR~7345. \label{tab_sfpobs}} 
	\tablehead{
		\colhead{\textbf{MJD(1D)}} &
		\colhead{\textbf{B(m)}} &
		\colhead{\textbf{$\vec{\rho}$(mas)}} &
		\colhead{\textbf{$\vec{\theta}$(deg)}} &
		\colhead{\textbf{\textit{HJD(2D)}}} &
		\colhead{\textbf{$JY_{calc}$}} &
		\colhead{\textbf{$\rho$(mas)}} &
		\colhead{\textbf{$\rho_{err}$(mas)}} &
			\colhead{\textbf{$\theta$}} &
			\colhead{\textbf{$\theta_{err}$}}
	}
	\startdata
53890.267	&	270.1	&	73.57	&	4.98	&	             	&	            	&	        	&	       	&	        	&	         \\ 
53891.208	&	275.73	&	73.15	&	132.82	&	             	&	            	&	        	&	       	&	        	&	         \\ 
53892.251	&	279.31	&	30.5	&	230.15	&	             	&	            	&	        	&	       	&	        	&	         \\ 
53892.28	&	297.19	&	37.11	&	226.82	&	             	&	            	&	        	&	       	&	        	&	         \\ 
53892.384	&	328.03	&	65.04	&	210.66	&	2453892.59802	&	2006.42538	&	63.13	&	0.9	&	341.1	&	  0.82   \\
\hline
53987.253	&	330.01	&	34.97	&	2.58	&	             	&	            	&	        	&	       	&	        	&	         \\
53988.221	&	330.29	&	41.28	&	9.42	&	             	&	            	&	        	&	       	&	        	&	         \\
53989.2249	&	271.08	&	64.26	&	308.46	&	             	&	            	&	        	&	       	&	        	&	         \\
53989.3145	&	227.63	&	47.34	&	294.65	&	             	&	            	&	        	&	       	&	        	&	         \\
53990.242	&	265.28	&	76.99	&	305.13	&	             	&	            	&	        	&	       	&	        	&	         \\
53991.1785	&	330.65	&	85.27	&	17.32	&	             	&	            	&	        	&	       	&	        	&	         \\
53991.2655	&	330.02	&	68.84	&	357.1	&	 2453989.74113\tablenotemark{1} 	&	2006.69609	&	87.76	&	7.05	&	352.08	&	  4.58   \\
\hline
54249.4004	&	276.69	&	61.62	&	24.1	&	             	&	            	&	        	&	       	&	        	&	         \\
54249.4071	&	277.11	&	61.02	&	22.84	&	             	&	            	&	        	&	       	&	        	&	         \\
54254.3993	&	177.45	&	80	&	346.58	&	             	&	            	&	        	&	       	&	        	&	         \\
54254.4941	&	174.47	&	78.8	&	327.75	&	2454251.92110	&	2007.41252	&	79.62	&	3.74	&	344.81	&	  2.69   \\
\hline
54323.2745	&	248.13	&	87.12	&	11.82	&	             	&	            	&	        	&	       	&	        	&	         \\
54324.3997	&	248.03	&	91.14	&	343.41	&	             	&	            	&	        	&	       	&	        	&	         \\
54329.1695	&	275.11	&	88.37	&	335.67	&	             	&	            	&	        	&	       	&	        	&	         \\
54330.4124	&	330.63	&	89.98	&	339.91	&	2454327.00411	&	2007.61756	&	92.88	&	1.53	&	351.32	&	  0.93   \\
\hline
54393.1171	&	271.57	&	53.78	&	308.82	&	             	&	            	&	        	&	       	&	        	&	         \\
54393.1858	&	242.88	&	42	&	297.83	&	             	&	            	&	        	&	       	&	        	&	         \\
54393.231	&	330.66	&	79.77	&	341.94	&	             	&	            	&	        	&	       	&	        	&	         \\
54393.2411	&	330.62	&	78.29	&	339.77	&	2454393.69400	&	2007.79930	&	78.62	&	4.71	&	359.5	&	  3.43   \\
\hline
54697.2614	&	248.12	&	84.4	&	9.66	&	             	&	            	&	        	&	       	&	        	&	         \\
54697.3486	&	248.12	&	85.59	&	350.19	&	             	&	            	&	        	&	       	&	        	&	         \\
54699.3158	&	248.06	&	85.69	&	356.1	&	             	&	            	&	        	&	       	&	        	&	         \\
54699.3208	&	248.07	&	86.4	&	354.9	&	             	&	            	&	        	&	       	&	        	&	         \\
54699.3942	&	247.63	&	82.28	&	339.01	&	             	&	            	&	        	&	       	&	        	&	         \\
54704.1787	&	247.29	&	76.88	&	337	&	             	&	            	&	        	&	       	&	        	&	         \\
54704.2825	&	248.38	&	86.94	&	359.39	&	2454701.77290	&	2008.63972	&	87.25	&	0.73	&	355.42	&	  0.48   \\
\hline
55115.1294	&	307.13	&	32.67	&	61.9	&	             	&	            	&	        	&	       	&	        	&	         \\
55115.1631	&	265.04	&	23.91	&	305.02	&	             	&	            	&	        	&	       	&	        	&	         \\
55115.1839	&	330.00	&	52.28	&	358.3	&	2455115.65951	&	2009.77592	&	53.6	&	0.63	&	8.99	&	  0.67   \\
\hline
55346.3611	&	272.87	&	82.92	&	342.19	&	             	&	            	&	        	&	       	&	        	&	         \\
55346.4009	&	327.78	&	74.37	&	31.08	&	             	&	            	&	        	&	       	&	        	&	         \\
55346.4467	&	313.37	&	18.52	&	73.45	&	2455347.38584	&	2010.40904	&	87.16	&	2.73	&	356.75	&	  1.79   \\
\hline
55439.17945	&	278.24	&	44.14	&	324.48	&	             	&	            	&	        	&	       	&	        	&	         \\
55439.1809	&	312.46	&	22.23	&	75.75	&	             	&	            	&	        	&	       	&	        	&	         \\
55439.1845	&	330.32	&	56.58	&	23.76	&	             	&	            	&	        	&	       	&	        	&	         \\
55439.2244	&	330.6	&	57.97	&	15.07	&	             	&	            	&	        	&	       	&	        	&	         \\
55439.2288	&	276.15	&	35.1	&	313.42	&	2455439.70121	&	2010.66310	&	59.17	&	0.86	&	7.13	&	  0.84   \\
\hline
55445.1374	&	246.68	&	53.66	&	25.48	&	             	&	            	&	        	&	       	&	        	&	         \\
55445.1407	&	154.22	&	19.31	&	70.45	&	             	&	            	&	        	&	       	&	        	&	         \\
55445.1439	&	177.44	&	50.82	&	345.69	&	2455445.64292	&	2010.67937	&	55.2	&	3.57	&	2.86	&	  3.70   \\
\hline
55468.1399	&	248.13	&	36.01	&	11.96	&	             	&	            	&	        	&	       	&	        	&	         \\
55468.1434	&	154.95	&	28.83	&	56.14	&	             	&	            	&	        	&	       	&	        	&	        \\
55465.147	&	177.18	&	26.15	&	338.59	&	 2455468.64471\tablenotemark{1} 		&	2010.74234	&	35.82	&	7.82	&	20.68	&	 12.39   \\
\hline
55516.1213	&	237.00	&	32.69	&	296.5	&	             	&	            	&	        	&	       	&	        	&	         \\
55516.1258	&	266.79	&	14.7	&	213.59	&	             	&	            	&	        	&	       	&	        	&	         \\
55516.1316	&	330.46	&	38.46	&	167.71	&	2455516.62513	&	2010.87372	&	40.59	&	0.25	&	330.4	&	  0.34   \\
\hline
55750.3251	&	278.27	&	55.3	&	324.29	&	             	&	            	&	        	&	       	&	        	&	         \\
55750.3276	&	312.38	&	21.45	&	75.88	&	             	&	            	&	        	&	       	&	        	&	         \\
55750.3312	&	330.26	&	66.13	&	24.07	&	2455750.83091	&	2011.51494	&	70.95	&	0.32	&	3.32	&	  0.76   \\
\hline
55775.2151	&	275.49	&	47.36	&	334.38	&	             	&	            	&	        	&	       	&	        	&	         \\
55775.2183	&	210.96	&	54.44	&	352.11	&	             	&	            	&	        	&	       	&	        	&	         \\
55775.2293	&	304.58	&	14.4	&	82.47	&	             	&	            	&	        	&	       	&	        	&	         \\
55775.2325	&	328.3	&	52.0	&	30.17	&	2455775.72680	&	2011.58309	&	56.42	&	0.04	&	7.28	&	  0.04   \\
\hline
55781.1517	&	177.33	&	51.55	&	1.96	&	             	&	            	&	        	&	       	&	        	&	         \\
55781.159	&	239.99	&	45.58	&	35.64	&	             	&	            	&	        	&	       	&	        	&	         \\
55781.3067	&	175.1	&	38.37	&	329.13	&	             	&	            	&	        	&	       	&	        	&	         \\
55781.3119	&	152.12	&	39.23	&	49.8	&	             	&	            	&	        	&	       	&	        	&	         \\
55781.3149	&	248.07	&	50.56	&	4.73	&	2455781.74181	&	2011.59958	&	51.1	&	0.9	&	9.83	&	  0.99   \\
\hline
55782.3172	&	270.81	&	23.86	&	308.27	&	             	&	            	&	        	&	       	&	        	&	         \\
55782.3209	&	301.25	&	35.83	&	57.27	&	             	&	            	&	        	&	       	&	        	&	         \\
55782.3248	&	330.13	&	51.01	&	6.04	&	 2455782.82393\tablenotemark{1} 		&	2011.60251	&	51.49	&	4.38	&	10.97	&	  4.96   \\
\hline
55796.15	&	274.78	&	29.46	&	336.45	&	             	&	            	&	        	&	       	&	        	&	         \\
55796.1535	&	296.71	&	14.21	&	86.26	&	             	&	            	&	        	&	       	&	        	&	         \\
55796.157	&	326.09	&	36.43	&	33.43	&	             	&	            	&	        	&	       	&	        	&	         \\
55796.1897	&	277.83	&	25.73	&	326.62	&	             	&	            	&	        	&	       	&	        	&	         \\
55796.193	&	310.74	&	17.28	&	77.96	&	             	&	            	&	        	&	       	&	        	&	         \\
55796.1961	&	329.88	&	36.73	&	25.94	&	2455796.67591	&	2011.64044	&	38.14	&	1.37	&	15.91	&	  2.08   \\
\hline
55808.1349	&	246.71	&	15.2	&	328.22	&	             	&	            	&	        	&	       	&	        	&	         \\
55808.1388	&	243.29	&	9.31	&	86.92	&	             	&	            	&	        	&	       	&	        	&	         \\
55808.1421	&	246.61	&	25.03	&	25.72	&	             	&	            	&	        	&	       	&	        	&	         \\
55808.2781	&	231.83	&	20.17	&	54.32	&	             	&	            	&	        	&	       	&	        	&	         \\
55808.2811	&	248.07	&	22.37	&	355.81	&	 2455808.6765\tablenotemark{1} 	&	2011.67336	&	19.7	&	11.06	&	24.68	&	  4.77   \\
\hline
55817.1117	&	276.43	&	 $\lesssim$5.00 	&	331.59	&	             	&	            	&	        	&	       	&	        	&	         \\
55817.1169	&	244.09	&	15.58	&	86.37	&	             	&	            	&	        	&	       	&	        	&	         \\
55817.1212	&	277.43	&	13.29	&	21.75	&	 2455817.61864\tablenotemark{1} 	&	2011.69778	&	17.2	&	7.63	&	61.45	&	 22.89   \\
\hline
55829.2295	&	256.35	&	18.48	&	301.6	&	             	&	            	&	        	&	       	&	        	&	         \\
55829.2337	&	283.33	&	2.92	&	45.38	&	             	&	            	&	        	&	       	&	        	&	         \\
55829.2378	&	330.05	&	11.98	&	356.26	&	             	&	            	&	        	&	       	&	        	&	         \\
55829.2602	&	239.47	&	17.07	&	297.04	&	             	&	            	&	        	&	       	&	        	&	         \\
55829.2669	&	267.39	&	5.94	&	34.07	&	             	&	            	&	        	&	       	&	        	&	         \\
55829.2709	&	330.41	&	13.62	&	348.44	&	2455829.74792	&	2011.73100	&	17.5	&	2.84	&	319.45	&	  9.30   \\
\hline
55843.1036	&	277.97	&	34.88	&	317.28	&	             	&	            	&	        	&	       	&	        	&	         \\
55843.1119	&	330.64	&	24.24	&	16.63	&	             	&	            	&	        	&	       	&	        	&	         \\
55843.1961	&	253.91	&	29.55	&	300.81	&	             	&	            	&	        	&	       	&	        	&	         \\
55843.2022	&	280.49	&	11.42	&	43.5	&	             	&	            	&	        	&	       	&	        	&	         \\
55843.2076	&	330.11	&	30.77	&	354.35	&	2455843.66510	&	2011.76910	&	34.68	&	1.48	&	331.39	&	  2.44   \\
\hline
56052.5008	&	278.33	&	65.96	&	143.79	&	             	&	            	&	        	&	       	&	        	&	         \\
56053.4454	&	244.54	&	70.32	&	210.35	&	             	&	            	&	        	&	       	&	        	&	         \\
56053.4514	&	150.23	&	21.55	&	255.33	&	             	&	            	&	        	&	       	&	        	&	         \\
56053.4546	&	177.4	&	80.97	&	170.95	&	             	&	            	&	        	&	       	&	        	&	         \\
56053.473	&	246.75	&	72.82	&	205.25	&	             	&	            	&	        	&	       	&	        	&	         \\
56053.4812	&	154.5	&	28.4	&	249.94	&	2456053.80120	&	2012.34442	&	81.5	&	0.99	&	359.19	&	  0.71   \\
\hline
56076.4242	&	177.43	&	67.75	&	163.86	&	             	&	            	&	        	&	       	&	        	&	         \\
56076.4289	&	155.64	&	37.11	&	247.25	&	             	&	            	&	        	&	       	&	        	&	         \\
56076.4307	&	247.64	&	68.06	&	159.02	&	             	&	            	&	        	&	       	&	        	&	         \\
56076.485	&	154.27	&	44.87	&	234.35	&	             	&	            	&	        	&	       	&	        	&	         \\
56076.4897	&	175.7	&	61.12	&	150.74	&	             	&	            	&	        	&	       	&	        	&	         \\
56076.4941	&	248.1	&	69.31	&	187.69	&	2456076.96660	&	2012.40785	&	71.25	&	1.54	&	2.73	&	  1.24   \\
\hline
56077.392	&	275.79	&	62.94	&	153.5	&	             	&	            	&	        	&	       	&	        	&	         \\
56077.3944	&	302.16	&	13.24	&	263.76	&	             	&	            	&	        	&	       	&	        	&	         \\
56077.3992	&	327.66	&	62.83	&	211.28	&	             	&	            	&	        	&	       	&	        	&	         \\
56077.4767	&	248.13	&	71.33	&	191.19	&	             	&	            	&	        	&	       	&	        	&	         \\
56077.4793	&	154.48	&	45.64	&	234.88	&	             	&	            	&	        	&	       	&	        	&	         \\
56077.4846	&	175.83	&	62.52	&	151.16	&	2456077.93962	&	2012.41051	&	72.49	&	1.56	&	3.45	&	  1.24   \\
\hline
56116.24469	&	177.32	&	49.3	&	179.69	&	             	&	            	&	        	&	       	&	        	&	         \\
56116.24826	&	202.91	&	18.56	&	259.21	&	             	&	            	&	        	&	       	&	        	&	         \\
56116.25173	&	291.68	&	41.7	&	221.66	&	             	&	            	&	        	&	       	&	        	&	         \\
56116.42285	&	236.22	&	16.74	&	120.8	&	             	&	            	&	        	&	       	&	        	&	         \\
56116.42728	&	234.39	&	36.53	&	236.21	&	             	&	            	&	        	&	       	&	        	&	         \\
56116.43033	&	248.05	&	47.59	&	177.37	&	2456116.84052	&	2012.51700	&	49.66	&	1.45	&	11.36	&	  1.66   \\
\hline
56131.18872	&	271.16	&	31.06	&	168.52	&	             	&	            	&	        	&	       	&	        	&	         \\
56131.19272	&	271.42	&	29.56	&	167.41	&	             	&	            	&	        	&	       	&	        	&	         \\
56131.19652	&	270.98	&	15.41	&	275.00	&	             	&	            	&	        	&	       	&	        	&	         \\
56131.20019	&	317.43	&	31.47	&	219.96	&	             	&	            	&	        	&	       	&	        	&	         \\
56131.23108	&	274.64	&	26.79	&	156.85	&	             	&	            	&	        	&	       	&	        	&	         \\
56131.23569	&	296.49	&	14.85	&	266.35	&	             	&	            	&	        	&	       	&	        	&	         \\
56131.23937	&	326.12	&	33.89	&	213.39	&	             	&	            	&	        	&	       	&	        	&	         \\
56132.15739	&	238.62	&	29.95	&	173.84	&	             	&	            	&	        	&	       	&	        	&	         \\
56132.16195	&	192.73	&	 $\lesssim$5.00 	&	288.72	&	             	&	            	&	        	&	       	&	        	&	         \\
56132.16563	&	232.54	&	31.61	&	220.4	&	             	&	            	&	        	&	       	&	        	&	         \\
56132.35159	&	174.54	&	32.58	&	147.89	&	             	&	            	&	        	&	       	&	        	&	         \\
56132.35758	&	151.00	&	32.26	&	227.72	&	             	&	            	&	        	&	       	&	        	&	         \\
56132.36157	&	248.06	&	31.82	&	183.19	&	2456132.18418	&	2012.55901	&	34.76	&	2.15	&	20.67	&	  3.55   \\
\hline
56141.20961	&	297.08	&	12.77	&	266.4	&	             	&	            	&	        	&	       	&	        	&	         \\
56141.21415	&	297.9	&	23.3	&	216.56	&	             	&	            	&	        	&	       	&	        	&	         \\
56141.35505	&	236.15	&	 $\lesssim$5.00 	&	120.76	&	             	&	            	&	        	&	       	&	        	&	         \\
56141.3618	&	291.29	&	20.82	&	230.56	&	             	&	            	&	        	&	       	&	        	&	         \\
56141.36727	&	300.95	&	21.98	&	183.13	&	2456141.80458	&	2012.58535	&	23.54	&	0.9	&	27.65	&	  2.17   \\
\hline
56148.19281	&	275.35	&	 $\lesssim$5.00 	&	154.79	&	             	&	            	&	        	&	       	&	        	&	         \\
56148.2227	&	309.19	&	12.67	&	259.37	&	             	&	            	&	        	&	       	&	        	&	         \\
56148.22548	&	329.55	&	12.03	&	207.15	&	2456148.72519	&	2012.60428	&	13.9	&	1.46	&	47.66	&	  6.04   \\
\hline
56162.25105	&	248.12	&	13.27	&	9.45	&	             	&	            	&	        	&	       	&	        	&	         \\
56162.25653	&	153.64	&	 $\lesssim$5.00 	&	52.9	&	             	&	            	&	        	&	       	&	        	&	         \\
56162.2631	&	175.13	&	22.99	&	329.2	&	             	&	            	&	        	&	       	&	        	&	         \\
56162.27947	&	248.06	&	15.21	&	3.05	&	             	&	            	&	        	&	       	&	        	&	         \\
56162.28356	&	149.85	&	 $\lesssim$5.00 	&	45.68	&	             	&	            	&	        	&	       	&	        	&	         \\
56162.28723	&	172.91	&	24.9	&	325.11	&	2456162.77546	&	2012.64276	&	23.1	&	2.5	&	319.35	&	  6.22   \\
\hline
56186.24046	&	261.4	&	38.24	&	303.44	&	             	&	            	&	        	&	       	&	        	&	         \\
56186.24455	&	289.15	&	13.98	&	49.16	&	             	&	            	&	        	&	       	&	        	&	         \\
56186.24988	&	330.00	&	40.7	&	358.69	&	             	&	            	&	        	&	       	&	        	&	         \\
56186.28818	&	235.69	&	37.15	&	296.23	&	             	&	            	&	        	&	       	&	        	&	         \\
56186.29246	&	266.21	&	24.64	&	33.13	&	             	&	            	&	        	&	       	&	        	&	         \\
56186.2966	&	330.45	&	46.66	&	347.72	&	2456186.77049	&	2012.70847	&	46.31	&	2.17	&	335.53	&	  2.69   \\
\hline
56195.1479	&	277.37	&	47.62	&	315.61	&	             	&	            	&	        	&	       	&	        	&	         \\
56195.15299	&	311.45	&	 $\lesssim$5.00 	&	246.52	&	             	&	            	&	        	&	       	&	        	&	         \\
56195.15704	&	330.57	&	43.47	&	14.62	&	             	&	            	&	        	&	       	&	        	&	         \\
56195.24809	&	245.41	&	40.82	&	298.45	&	             	&	            	&	        	&	       	&	        	&	         \\
56195.25151	&	273.78	&	22.71	&	38.36	&	             	&	            	&	        	&	       	&	        	&	         \\
56195.25502	&	320.24	&	49.71	&	351.64	&	2456195.70359	&	2012.73293	&	51.89	&	2.35	&	337.47	&	  2.59   \\
\hline
56245.07452	&	263.2	&	55.46	&	304.19	&	             	&	            	&	        	&	       	&	        	&	         \\
56245.08012	&	290.44	&	32.53	&	50	&	             	&	            	&	        	&	       	&	        	&	         \\
56245.08698	&	330.00	&	84.27	&	359.11	&	2456245.57954	&	2012.86949	&	81.08	&	7.34	&	346.98	&	  5.19   \\
\hline
56414.46069	&	210.95	&	70.75	&	353.93	&	             	&	            	&	        	&	       	&	        	&	         \\
56414.4654	&	214.64	&	30.2	&	72.75	&	             	&	            	&	        	&	       	&	        	&	         \\
56414.47028	&	326.88	&	57.76	&	32.43	&	             	&	            	&	        	&	       	&	        	&	         \\
56414.49626	&	210.94	&	73.2	&	346.07	&	             	&	            	&	        	&	       	&	        	&	         \\
56414.50139	&	220.89	&	36.77	&	65.72	&	             	&	            	&	        	&	       	&	        	&	         \\
56414.50395	&	329.91	&	56.75	&	25.84	&	2456414.98359	&	2013.33329	&	70.1	&	9.27	&	4.13	&	  7.57   \\
\hline
56438.36719	&	272.58	&	48.7	&	343.13	&	             	&	            	&	        	&	       	&	        	&	         \\
56438.37547	&	321.58	&	47.3	&	37.47	&	             	&	            	&	        	&	       	&	        	&	         \\
56438.41196	&	177.43	&	53.21	&	348.45	&	             	&	            	&	        	&	       	&	        	&	         \\
56438.41639	&	153.49	&	29.56	&	71.58	&	             	&	            	&	        	&	       	&	        	&	         \\
56438.42247	&	246.92	&	54.51	&	24.64	&	2456438.90039	&	2013.39877	&	55.76	&	2.99	&	11.31	&	  3.07   \\
\hline
56445.39279	&	177.43	&	46.9	&	348.5	&	             	&	            	&	        	&	       	&	        	&	         \\
56445.40031	&	246.79	&	49.29	&	25.11	&	             	&	            	&	        	&	       	&	        	&	         \\
56445.42323	&	278.3	&	32.64	&	324.1	&	             	&	            	&	        	&	       	&	        	&	         \\
56445.42937	&	312.85	&	23.86	&	74.99	&	             	&	            	&	        	&	       	&	        	&	         \\
56445.43394	&	330.41	&	48.72	&	23.04	&	2456445.91792	&	2013.41798	&	50.35	&	1.14	&	12.96	&	  1.83   \\
\hline
56461.31538	&	273.61	&	30.52	&	339.9	&	             	&	            	&	        	&	       	&	        	&	         \\
56461.32094	&	290.43	&	9.38	&	88.7	&	             	&	            	&	        	&	       	&	        	&	         \\
56461.32531	&	324.35	&	34.8	&	35.24	&	2456461.82305	&	2013.46152	&	37.07	&	0.46	&	13.78	&	  0.71   \\
\hline
56518.21606	&	278.04	&	32.28	&	198.81	&	             	&	            	&	        	&	       	&	        	&	         \\
56518.21948	&	249.3	&	13.36	&	181.51	&	             	&	            	&	        	&	       	&	        	&	         \\
56518.22234	&	278.28	&	46.42	&	144.22	&	             	&	            	&	        	&	       	&	        	&	         \\
56518.28757	&	278.76	&	40.09	&	183.73	&	             	&	            	&	        	&	       	&	        	&	         \\
56518.29165	&	245.37	&	 $\lesssim$5.00 	&	245.26	&	             	&	            	&	        	&	       	&	        	&	         \\
56518.29646	&	272.05	&	42.04	&	129.18	&	2456518.75849	&	2013.61740	&	46.33	&	1.01	&	334.42	&	  1.26   \\
\hline
56566.1015	&	156.22	&	19.83	&	244.27	&	             	&	            	&	        	&	       	&	        	&	         \\
56566.10548	&	210.68	&	72.2	&	160.99	&	             	&	            	&	        	&	       	&	        	&	         \\
56566.10922	&	278.55	&	63.22	&	194.06	&	             	&	            	&	        	&	       	&	        	&	         \\
56566.24919	&	235.06	&	49.7	&	116.1	&	             	&	            	&	        	&	       	&	        	&	         \\
56566.25368	&	210.1	&	45.29	&	142.36	&	             	&	            	&	        	&	       	&	        	&	         \\
56566.25728	&	278.2	&	73.17	&	162.28	&	2456566.68059	&	2013.74861	&	73.3	&	4.04	&	346.7	&	  3.17   \\
\hline
56772.48997	&	245.74	&	53.27	&	208.03	&	             	&	            	&	        	&	       	&	        	&	         \\
56772.49743	&	177.43	&	50.56	&	168.39	&	             	&	            	&	        	&	       	&	        	&	         \\
56772.50785	&	246.9	&	51.62	&	204.72	&	             	&	            	&	        	&	       	&	        	&	         \\
56772.50972	&	154.64	&	26.19	&	249.68	&	             	&	            	&	        	&	       	&	        	&	         \\
56772.51533	&	177.44	&	49.05	&	164.53	&	             	&	            	&	        	&	       	&	        	&	         \\
56772.52139	&	155.6	&	30.73	&	247.37	&	2456773.00713	&	2014.31351	&	54.51	&	1.89	&	10.14	&	  1.98   \\
\hline
56826.36376	&	248.13	&	19.83	&	324.12	&	             	&	            	&	        	&	       	&	        	&	         \\
56827.3513	&	247.3	&	20.15	&	326.82	&	             	&	            	&	        	&	       	&	        	&	         \\
56827.35608	&	307.03	&	12.33	&	260.96	&	             	&	            	&	        	&	       	&	        	&	         \\
56827.44708	&	244.36	&	22.25	&	306.09	&	             	&	            	&	        	&	       	&	        	&	         \\
56827.46087	&	301.39	&	10.52	&	10.15	&	2456827.69816	&	2014.46323	&	20.91	&	1.43	&	311.81	&	  3.95   \\
\hline
56935.14887	&	273.59	&	67.45	&	310.48	&	             	&	            	&	        	&	       	&	        	&	         \\
56935.15368	&	304.99	&	26.97	&	60.1	&	             	&	            	&	        	&	       	&	        	&	         \\
56935.15879	&	330.24	&	71.62	&	8.41	&	2456935.65478	&	2014.75881	&	80.56	&	5.34	&	347.28	&	  3.78   \\
\hline
57566.26909	&	271.84	&	79.94	&	345.73	&	             	&	            	&	        	&	       	&	        	&	         \\
57566.27375	&	275.62	&	27.12	&	273.61	&	             	&	            	&	        	&	       	&	        	&	         \\
57566.27962	&	319.97	&	48.68	&	38.53	&	2457566.77685	&	2016.48672	&	80.83	&	1.53	&	344.65	&	  1.09   \\
\hline
57573.31056	&	210.97	&	82.41	&	348.99	&	             	&	            	&	        	&	       	&	        	&	         \\
57573.31825	&	219.78	&	18.99	&	67.69	&	             	&	            	&	        	&	       	&	        	&	         \\
57573.32459	&	329.46	&	60.21	&	27.42	&	2457573.82070	&	2016.50600	&	80.21	&	4.33	&	350.87	&	  3.09   \\
\hline
57650.21089	&	173.26	&	79.92	&	325.63	&	             	&	            	&	        	&	       	&	        	&	         \\
57650.21478	&	149.01	&	60.47	&	44.23	&	             	&	            	&	        	&	       	&	        	&	         \\
57650.21898	&	248.05	&	88.95	&	0.17	&	             	&	            	&	        	&	       	&	        	&	         \\
57650.27145	&	162.76	&	72.15	&	316.64	&	             	&	            	&	        	&	       	&	        	&	         \\
57650.27702	&	138.5	&	77.35	&	24.46	&	             	&	            	&	        	&	       	&	        	&	         \\
57650.28224	&	248.11	&	77.84	&	346.27	&	2457650.74032	&	2016.71661	&	90.61	&	2.08	&	354.54	&	  1.32   \\
\hline
&		   &	   &		& 2458143.06401 &	2018.06453	&	3.4	&	0.04	&	199.82	&	  0.59  \\				
\enddata
	\tablecomments{All CHARA Array SFP observations for HR~7345.  
Each set of 1-D vector observations (along with the projected baseline length 
and epoch of observation) in columns one through four were combined to 
create the true location of the secondary and average time of all the 
data points defined in the last six columns with position angle being defined as standard North through East without correcting for precession. The MJD from the 1-D measurements was converted to HJD-2400000 to match the time coordinates of the spectroscopic data included in \ref{tab_SBobs}. Additionally, we omitted the measurement from HJD~56478.4694 (sep=25.57, 
PA=131.04) because of the very large discrepancy with the orbital fit and 
because it was measured from three points on only one baseline and was 
near periastron. The final line of the table is the measurement taken by three-baseline CLIMB during periastron in January 2018 and does not include 1-D information.}

\tablenotetext{1}{The residuals from these measurements compared with the orbit fit were more than three times the measurement error, so we increased their uncertainties by a factor of 10 to minimize their impact on the orbit fit.}
\end{deluxetable}

\clearpage

\begin{deluxetable}{rrrrr}
	\tablewidth{0pt}
	\tabletypesize{\footnotesize}
	\tablecaption{Spectroscopic Observations for system HR~7345. \label{tab_SBobs}} 
	\tablehead{
		\colhead{\textbf{HJD - 2400000}} &
		\colhead{\textbf{$K_{1}$(km/s)}} &
		\colhead{\textbf{$K_{1}$ weight}} &
		\colhead{\textbf{$K_{2}$(km/s)}} &
		\colhead{\textbf{$K_{2}$ weight}}
}
\startdata
51853.2540 &  -4.51 & 1.0 &  8.49 & 1.0  \\
\hline
56822.6593 &  -8.57 & 1.0 & 12.52 & 1.0  \\   
56835.9192 &  -1.74 & 0.0 &  6.19 & 0.0  \\   
56898.7493 &   2.07 & 0.0 &       &      \\
56934.7889 &   2.11 & 0.0 &       &      \\
56979.6116 &   2.06 & 0.0 &       &      \\
\hline
57047.0342 &   2.21 & 0.0 &       &      \\
57080.9864 &   2.03 & 0.0 &       &      \\
57120.8380 &   2.11 & 0.0 &       &      \\
57164.8253 &  -2.67 & 0.0 &  6.82 & 0.0  \\   
57282.6075 &   1.96 & 0.0 &       &      \\
57283.6763 &   2.00 & 0.0 &       &      \\
\hline
57417.0219 &   1.99 & 0.0 &       &      \\
57432.0365 &   1.74 & 0.0 &       &      \\
57451.9356 &   1.67 & 0.0 &       &      \\
57474.8658 &  -4.63 & 0.5 &  8.21 & 0.5  \\   
57487.8338 &  -6.75 & 0.5 & 10.13 & 0.5  \\   
57497.8088 &  -2.57 & 0.0 &  5.88 & 0.0  \\   
57500.9954 &  -1.87 & 0.0 &  5.68 & 0.0  \\   
57509.7723 &  -0.51 & 0.0 &  4.48 & 0.0  \\   
57514.8204 &   1.92 & 0.0 &       &      \\
57515.8037 &   1.94 & 0.0 &       &      \\
57516.8238 &   1.94 & 0.0 &       &      \\
57517.7608 &   2.10 & 0.0 &       &      \\
57524.7618 &   2.00 & 0.0 &       &      \\
57527.7524 &   1.89 & 0.0 &       &      \\
57530.7197 &   1.91 & 0.0 &       &      \\
57535.9460 &   1.86 & 0.0 &       &      \\
57538.7014 &   1.78 & 0.0 &       &      \\
57580.7087 &   1.75 & 0.0 &       &      \\
57629.8280 &   2.07 & 0.0 &       &      \\
57678.7602 &   2.06 & 0.0 &       &      \\
57716.6133 &   1.96 & 0.0 &       &      \\
57717.6021 &   2.14 & 0.0 &       &      \\
57722.6474 &   2.06 & 0.0 &       &      \\
57748.5772 &   2.00 & 0.0 &       &      \\
\hline
57761.0544 &   1.91 & 0.0 &       &      \\
57762.0435 &   1.88 & 0.0 &       &      \\
57763.0429 &   1.79 & 0.0 &       &      \\
57765.0411 &   1.92 & 0.0 &       &      \\
57777.0446 &   1.71 & 0.0 &       &      \\
57779.0066 &   2.10 & 0.0 &       &      \\
57784.9812 &   2.04 & 0.0 &       &      \\
57787.9734 &   1.69 & 0.0 &       &      \\
57791.9718 &   1.73 & 0.0 &       &      \\
57794.9856 &   1.87 & 0.0 &       &      \\
57796.0239 &   1.73 & 0.0 &       &      \\
57800.9306 &   1.82 & 0.0 &       &      \\
57806.9195 &  -5.22 & 0.5 &  9.58 & 0.5  \\   
57807.0310 &  -5.49 & 0.5 &  9.78 & 0.5  \\   
57807.9270 &  -8.13 & 1.0 & 12.20 & 1.0  \\   
57808.9264 & -13.32 & 1.0 & 17.70 & 1.0  \\   
57809.0310 & -14.38 & 1.0 & 18.45 & 1.0  \\   
57809.9104 & -24.74 & 1.0 & 28.24 & 1.0  \\   
57809.9410 & -25.10 & 1.0 & 28.89 & 1.0  \\   
57809.9810 & -25.59 & 1.0 & 29.64 & 1.0  \\      
57810.0010 & -25.92 & 1.0 & 30.04 & 1.0  \\   
57810.0261 & -26.23 & 1.0 & 30.62 & 1.0  \\   
57812.0272 & -36.62 & 1.0 & 40.93 & 1.0  \\   
57813.9412 & -18.22 & 1.0 & 21.99 & 1.0  \\   
57813.9486 & -18.14 & 1.0 & 21.91 & 1.0  \\   
57813.9612 & -18.16 & 1.0 & 21.61 & 1.0  \\   
57813.9812 & -17.91 & 1.0 & 21.82 & 1.0  \\   
57814.0012 & -17.87 & 1.0 & 21.64 & 1.0  \\   
57814.0262 & -17.65 & 1.0 & 21.46 & 1.0  \\   
57814.8998 & -14.53 & 1.0 & 18.01 & 1.0  \\   
57814.9612 & -14.29 & 1.0 & 17.77 & 1.0  \\   
57816.8990 &  -9.81 & 1.0 & 13.04 & 1.0  \\   
57816.9613 &  -9.54 & 1.0 & 12.88 & 1.0  \\   
57818.9437 &  -7.16 & 1.0 & 10.55 & 1.0  \\   
57819.0214 &  -6.92 & 1.0 & 10.58 & 1.0  \\   
57819.8909 &  -6.16 & 1.0 & 10.00 & 1.0  \\   
57820.0164 &  -6.29 & 1.0 &  9.65 & 1.0  \\   
57820.9126 &  -5.58 & 1.0 &  9.25 & 1.0  \\   
57820.9614 &  -5.66 & 1.0 &  9.02 & 1.0  \\   
57821.9110 &  -5.13 & 1.0 &  8.60 & 1.0  \\   
57822.0164 &  -5.20 & 1.0 &  8.45 & 1.0  \\   
57822.8834 &  -4.41 & 0.5 &  8.30 & 0.5  \\   
57823.0165 &  -4.41 & 0.5 &  8.24 & 0.5  \\   
57823.8719 &  -4.43 & 0.5 &  7.54 & 0.5  \\   
57824.8732 &  -3.92 & 0.5 &  7.29 & 0.5  \\   
57825.8731 &  -3.78 & 0.0 &  6.69 & 0.0  \\   
57826.8731 &  -3.57 & 0.0 &  6.28 & 0.0  \\   
57827.8731 &  -3.22 & 0.0 &  6.40 & 0.0  \\   
57829.8737 &  -2.25 & 0.0 &  6.16 & 0.0  \\   
57830.8733 &  -2.21 & 0.0 &  5.72 & 0.0  \\   
57831.8751 &  -2.01 & 0.0 &  5.94 & 0.0  \\   
57832.8779 &  -1.90 & 0.0 &  5.64 & 0.0  \\   
57843.0072 &   2.00 & 0.0 &       &      \\
57863.8002 &   1.96 & 0.0 &       &      \\
\hline
58139.0144 &  -6.39 & 1.0 & 10.50 & 1.0  \\
58139.0305 &  -6.75 & 1.0 & 10.50 & 1.0  \\
58140.0218 &  -9.83 & 1.0 & 14.53 & 1.0  \\
58141.0305 &  -17.9 & 1.0 & 22.14 & 1.0  \\
58141.0379 & -17.87 & 1.0 & 22.26 & 1.0  \\
58142.0185 & -34.15 & 1.0 & 38.59 & 1.0  \\
58142.0257 & -34.08 & 1.0 & 38.88 & 1.0  \\
58142.0329 & -34.38 & 1.0 & 38.88 & 1.0  \\
58142.0403 & -34.53 & 1.0 & 39.20 & 1.0  \\
58142.0474 & -34.79 & 1.0 & 39.25 & 1.0  \\
58142.9958 & -46.30 & 1.0 & 50.47 & 1.0  \\
58143.0031 & -46.20 & 1.0 & 50.50 & 1.0  \\
58143.0185 & -46.10 & 1.0 & 50.44 & 1.0  \\
58143.0257 & -46.17 & 1.0 & 50.28 & 1.0  \\
58143.0329 & -46.11 & 1.0 & 50.16 & 1.0  \\
58144.0185 & -30.77 & 1.0 & 35.07 & 1.0  \\
58144.0257 & -30.99 & 1.0 & 34.96 & 1.0  \\
58144.0329 & -30.74 & 1.0 & 34.90 & 1.0  \\
58144.0402 & -30.42 & 1.0 & 34.90 & 1.0  \\
58144.0474 & -30.44 & 1.0 & 34.71 & 1.0  \\
58144.9979 & -21.14 & 1.0 & 25.36 & 1.0  \\
58145.0051 & -21.38 & 1.0 & 25.10 & 1.0  \\
58145.0185 & -21.50 & 1.0 & 24.95 & 1.0  \\
58145.0257 & -21.22 & 1.0 & 25.10 & 1.0  \\
58145.0329 & -21.21 & 1.0 & 24.88 & 1.0  \\
58146.0306 & -15.98 & 1.0 & 20.09 & 1.0  \\
58146.0415 & -16.00 & 1.0 & 19.78 & 1.0  \\
58147.0101 & -12.42 & 1.0 & 16.76 & 1.0  \\
58147.0415 & -12.34 & 1.0 & 16.56 & 1.0  \\
58148.0415 & -10.33 & 1.0 & 14.17 & 1.0  \\
58149.0041 &  -8.88 & 1.0 & 12.37 & 1.0  \\
58149.0415 &  -8.61 & 1.0 & 12.47 & 1.0  \\
58150.0119 &  -7.23 & 1.0 & 11.36 & 1.0  \\
58150.0416 &  -7.19 & 1.0 & 11.48 & 1.0  \\
58150.9796 &  -6.43 & 1.0 & 10.86 & 1.0  \\
58151.0416 &  -6.18 & 1.0 & 10.42 & 1.0  \\
58152.0160 &  -5.69 & 1.0 &  9.72 & 1.0  \\
58152.0416 &  -5.92 & 1.0 &  9.49 & 1.0  \\
58153.9725 &  -4.66 & 0.5 &  8.34 & 0.5  \\
58154.0416 &  -4.67 & 0.5 &  8.50 & 0.5  \\
58154.9681 &  -4.19 & 0.5 &  8.09 & 0.5  \\
58155.0416 &  -4.21 & 0.5 &  7.95 & 0.5  \\
\enddata
\tablecomments{Spectroscopic observations for HR~7345 obtained 
from 2014 to 2018 with the 2~m AST and a fiber-fed echelle spectrograph 
at Fairborn Observatory. The velocity from the single, partially 
resolved ELODIE spectrum acquired in 2000 November \citep{Prug07} is 
also listed.}
\end{deluxetable}

\clearpage

\begin{deluxetable}{lccc}
	\tabletypesize{\footnotesize}
	\tablecaption{Orbital Parameters for HR~7345. \label{orbittab}}
	\tablewidth{0pt}
	\tablehead{
		\colhead{Parameter} & \colhead{SB2 Orbit} & \colhead{VB Orbit} & \colhead{Joint Fit}  }
	\startdata
	$P$ (d)                  & 331.607   $\pm$ 0.0037  & 331.601  $\pm$ 0.075  & 331.609   $\pm$ 0.0037    \\
	$T$ (HJD)                & 58142.692 $\pm$ 0.0029  & 58142.681 $\pm$ 0.012   & 58142.690 $\pm$ 0.0027    \\
	$e$                      & 0.93209   $\pm$ 0.00013 & 0.9324   $\pm$ 0.0011 & 0.9322   $\pm$ 0.00013  \\
	$a$ (mas)                &                         & 47.58    $\pm$ 0.11   & 47.432    $\pm$ 0.035    \\
	$i$ ($^\circ$)            &                         & 29.6     $\pm$ 1.2    & 29.48    $\pm$ 0.86     \\
	$\Omega$ ($^\circ$)       &                         & 176.2    $\pm$ 6.4    & 181.046   $\pm$ 0.092     \\
	$\omega_{\rm A}$ ($^\circ$) & 169.934  $\pm$ 0.077  & 175.4    $\pm$ 7.3    & 169.888   $\pm$ 0.075     \\
	$K_A$ (km\,s$^{-1}$)     & 25.535    $\pm$ 0.047   &                       & 25.555    $\pm$ 0.047     \\
	$K_B$ (km\,s$^{-1}$)     & 25.927    $\pm$ 0.047   &                       & 25.947     $\pm$ 0.048     \\
	$\gamma$ (km\,s$^{-1}$)  & 1.827     $\pm$ 0.030   &                       & 1.827     $\pm$ 0.031    \\
	\enddata 
	\tablecomments{The angle between the ascending node and periastron, as referenced to HR~7345 B (the typical reference for visual orbits), is given by $\omega_{\rm B}$ = $\omega_{\rm A}$ + 180$^\circ$ = 349\fdg89 $\pm$ 0\fdg08.}
\end{deluxetable} 

\clearpage

\begin{deluxetable}{lc}
	\tabletypesize{\footnotesize}
	\tablecaption{Stellar Properties for HR~7345 \label{orbparam}}
	\tablewidth{0pt}
	\tablehead{
		\colhead{Parameter} & \colhead{Value}  }
	\startdata
	$M_{\rm A}$ ($M_\odot$)  & 0.941 $\pm$ 0.076 \\
	$M_{\rm B}$ ($M_\odot$)  & 0.926 $\pm$ 0.075 \\
	$d$ (pc)              & 24.34 $\pm$ 0.45 \\
	$\pi$ (mas)           & 41.08 $\pm$ 0.77 \\
\enddata 
\end{deluxetable}                                                                       

\clearpage

\clearpage

\begin{thebibliography}{}

\bibitem[Ambartsumian(1937)]{AMB1937}Ambartsumian, V.A., \ 1937, Astronomicheskii Zhurnal, 14, 207 

\bibitem[Anosova(1986)]{ANA1986} Anosova, J.~P.\ 1986, \apss, 124, 217 

\bibitem[Barker et al.(1967)]{betal67} Barker, E. S., Evans, D. S., \& Laing, J. D. 1967, RGOB, 130, 355

\bibitem[Barnes III et al.(1986)]{BA86} Barnes III, T.~G., Moffett, T.~J., Slovak, M.~H. \ 1986, \pasp, 98, 223

\bibitem[Bate(2009)]{BATE2009} Bate, M.~R.\ 2009, \mnras, 392, 590 

\bibitem[Beavers et al.(1979)]{BE79} Beavers, W.~I., Eitter, J.~J., Ketelsen, D.~A., \& Oesper, D.~A. \ 1979, \pasp, 91, 698

\bibitem[Bourges et al.(2014)]{JMMC} Bourges, L., Lafrasse, S., Mella, G., Chesneau, O., Bouquin, J.~L., Duvert, G., Chelli, A., Delfosse, X. \ 2014, Astronomical Society of the Pacific Conference Series, 485, 223

\bibitem[ten Brummelaar et al.(2005)]{CHARA2} ten Brummelaar, T.~A., McAlister, H.~A., Ridgway, S.~T., Bagnuolo, W.~G.~Jr., Turner, N.~H., Sturmann, L., Sturmann, J., Berger, D.~H., Ogden, C.~E., Cadman, R., Hartkopf, W.~I., Hopper, C.~H., \& Shure, M.~A.\ 2005, \apj, 628, 453

\bibitem[ten Brummelaar et al.(2008)]{CLIMB} ten Brummelaar, T.~A., et al.\ 2008, \procspie, 7013,  

\bibitem[Ten Brummelaar et al.(2013)]{CLIMB2} Ten Brummelaar, T.~A., Sturmann, J., Ridgway, S.~T., et al.\ 2013, Journal of Astronomical Instrumentation , 2, 1340004 

\bibitem[Crifo et al.(2010)]{Crif10} Crifo, F., Jasniewicz, G., Soubiran, C., et al.\ 2010, \aap, 524, A10 

\bibitem[Crissman(1957)]{CRIS57} Crissman, B.~G., \ 1957, \aj \phn 62, 280

\bibitem[Duflot et al.(1995)]{Duf95} Duflot, M., Figon, P., \& Meyssonnier, N.\ 1995, \aaps, 114, 269 

\bibitem[Duquennoy et al.(1991)]{DM1} Duquennoy, A., Mayor, M. Halbwachs, J.~L. \ 1991, \aaps, 88, 281 

\bibitem[Duquennoy \& Mayor(1991)]{DM2} Duquennoy, A., \& Mayor, M.\ 1991, \aap, 248, 485 

\bibitem[Eaton \& Williamson(2007)]{ew07} Eaton, J. A., \& Williamson, M. H. 2007, \pasp, 119, 886

\bibitem[Farrington et al.(2010)]{FAR2010} Farrington, C.~D., ten~Brummelaar, T.~A., Mason, B.~D., Hartkopf, W.~I., McAlister, H.~A., Raghavan, D., Turner, N.~H., Sturmann, L., Sturmann, J., \& Ridgway, S.~T., \ 2010, \aj \phn 139, 2308

\bibitem[Farrington et al.(2014)]{FAR2014} Farrington, C.~D., ten~Brummelaar, T.~A., Mason, B.~D., Hartkopf, W.~I., McAlister, H.~A., Mourard, D., Moravveji, E. Turner, N.~H., Sturmann, L., \& Sturmann, J., \ 2014, \aj \phn 148, 48

\bibitem[Fekel \& Griffin(2011)]{fg11} Fekel., F. C. \& Griffin, R. F. 2011, The Observatory, 131, 283

\bibitem[Fekel et al.(2009)]{ftw09} Fekel, F. C., Tomkin, J., \& Williamson, M. H. 2009, \aj, 137, 3900

\bibitem[Fehrenbach et al.(1997)]{Feh97} Fehrenbach, C., Duflot, M., Mannone, C., Burnage, R., \& Genty, V.\ 1997, \aaps, 124, 

\bibitem[Finsen(1936)]{FIN1936} Finsen, W.~S.\ 1936, \mnras, 96, 862

\bibitem[Gaia Collaboration et al.(2016)]{Gaia16} Gaia Collaboration, Brown, A.~G.~A., Vallenari, A., et al.\ 2016, \aap, 595, A2 

\bibitem[Goldin \& Makarov(2006)]{GOLD06} Goldin, A., \& Makarov, V.~V.\ 2006, \apjs, 166, 341 

\bibitem[Gray et al.(2003)]{Gray03} Gray, R.~O., Corbally, C.~J., Garrison, R.~F., McFadden, M.~T., \& Robinson, P.~E.\ 2003, \aj, 126, 2048 

\bibitem[Griffin(2003)]{GRI2} Griffin, R.~F.,\ 2003, The Observatory, 123, 344

\bibitem[Griffin(2012)]{GRI2012} Griffin, R.~F.\ 2012, Journal of Astrophysics and Astronomy, 33, 29 

\bibitem[Halliday(1955)]{HALI55} Halliday, I., \ 1955, \apj \phn 122, 222

\bibitem[Halbwachs et al.(2003)]{Halb03} Halbwachs, J.~L., Mayor, M., Udry, S., \& Arenou, F.\ 2003, \aap, 397, 159  

\bibitem[Hartkopf et al.(2001)]{6TH} Hartkopf, W.~I., Mason, B.~D., \& Worley, C.~E.\ 2001, \aj, 122, 3472 

\bibitem[Holmberg et al.(2009)]{Holm09} Holmberg, J., Nordstr{\"o}m, B., \& Andersen, J.\ 2009, \aap, 501, 941 

\bibitem[Hut(1981)]{h81} Hut, P. 1981, \aap, 99, 126

\bibitem[Hut(1983)]{h83} Hut, P.\ 1983, \apj, 268, 342 

\bibitem[Kluska et al.(2018)]{KLUSKA2018} Kluska, J., Kraus, S., Davies, C.~L., Harries, T., Willson, M., Monnier, J.~D., Aarnio, A., Baron, F., Millan-Gabet, R., Ten Brummelaar, T.~A., Che, X., Hinkley, S., Preibisch, T., Sturmann, J., Sturmann, L., Touhami, Y. \ 2018, \aj, 855, 44 

\bibitem[Kozai(1962)]{K62} Kozai, Y.\ 1962, \aj, 67, 591 

\bibitem[Lacy \& Fekel(2011)]{lf11} Lacy C. H. S., \& Fekel, F. C. 2011, \aj, 142, 185

\bibitem[Lidov(1962)]{L62} Lidov, M.~L.\ 1962, \planss, 9, 719 

\bibitem[Luri et al.(2018)]{LURI2018} Luri, X., Brown, A.~G.~A., Sarro, L.~M., et al.\ 2018, arXiv:1804.09376 

\bibitem[{\it The Washington Double Star Catalog}, Mason et al.(2001-2009)]{WDS} Mason, B.~D., Wycoff, G.~L., Hartkopf, W.~I., Douglass, G.~G., and Worley, C.~E., \ 2001, \aj, 122, 3466-3471 

\bibitem[McAlister et al.(1987)]{McA1987} McAlister, H.~A., Hartkopf, W.~I., Hutter, D.~J., Franz, O.~G., \ 1987, \aj, 91, 688

\bibitem[Monnier et al.(2011)]{MONNIER2011} Monnier, J.~D., Zhao, M., Pedretti, E., Millan-Gabet, R., Berger, J.~-P., Traub, W., Schloerb, F.~P., ten Brummelaar, T.~A., McAlister, H.~A., Ridgway, S., Sturmann, L., Sturmann, J., Turner, N., Baron, F., Kraus, S., Tannirkulam, A., Williams, P.~M., \ 2011, \apj, 742, L1 

\bibitem[Nidever et al.(2002)]{Nid02} Nidever, D.~L., Marcy, G.~W., Butler, R.~P., Fischer, D.~A., \& Vogt, S.~S.\ 2002, \apjs, 141, 503 

\bibitem[Nordstr{\"o}m et al.(2004)]{Nord} Nordstr{\"o}m, B., Mayor, M., Andersen, J.,  Holmberg, J., Pont, F., J{\o}rgensen, B.~R., Olsen, E.~H., Udry, S., and Mowlavi, N., \ 2004, \aap, 418, 989

\bibitem[Perryman et al.(1997)]{HIP1} Perryman, M.~A.~C., Lindegren, L., Kovalevsky, J., Hoeg, E., Bastian, U., Bernacca, P.~L., Cr{\'e}z{\'e}, M., Donati, F., Grenon, M., van Leeuwen, F., van der Marel, H., Mignard, F., Murray, C.~A., Le Poole, R.~S., chrijver, H., Turon, C., Arenou, F., Froeschl{\'e}, M., and Petersen, C.~S, \ 1997, \aap, 323, L49

\bibitem[Pourbaix et al.(2004)]{9TH} Pourbaix, D., Tokovinin, A.~A., Batten, A.~H., et al.\ 2004, \aap, 424, 727

\bibitem[Prugniel et al.(2007)]{Prug07} Prugniel, P., Soubiran, C., Koleva, M., \& Le Borgne, D.\ 2007, VizieR Online Data Catalog, 3251,  

\bibitem[Raghavan et al.(2010)]{RAG2010} Raghavan, D., McAlister, H.~A., Henry, T.~J., et al.\ 2010, \apjs, 190, 1 

\bibitem[Sahlmann et al.(2011)]{SAH} Sahlmann, J., S{\'e}gransan, D., Queloz, D., et al.\ 2011, \aap, 525, A95 

\bibitem[Scarfe(2010)]{s10} Scarfe, C. D. 2010, The Observatory, 130, 214

\bibitem[Schaefer et al.(2006)]{schaefer06} Schaefer, G.~H., Simon, M., Beck, T.~L., Nelan, E., \& Prato, L.\ 2006, \aj, 132, 2618 

\bibitem[Schaefer et al.(2016)]{schaefer16} Schaefer, G. H., Hummel, C. A., Gies, D. R., et al. 2016, \aj, 152, 213


\bibitem[Soubiran et al.(2013)]{Soub13} Soubiran, C., Jasniewicz, G., Chemin, L., et al.\ 2013, \aap, 552, A64 

\bibitem[Strassmeier et al.(2013)]{STRA} Strassmeier, K.~G., Weber, M., \& Granzer, T.\ 2013, \aap, 559, A17 

\bibitem[Tokovinin(2016)]{Toko2016} Tokovinin, A.\ 2016, \aj, 152, 138 

\bibitem[Tokovinin \& Kiyaeva(2016)]{TOKO2016b} Tokovinin, A., \& Kiyaeva, O.\ 2016, \mnras, 456, 2070 

\bibitem[Torres et al.(2010)]{Torres10}  
Torres, G., Andersen, J., \& Gim\'enez, A. 2010, Astronomy \& Astrophysics Review, 18, 67

\bibitem[Valenti \& Fischer(2005)]{Val05} Valenti, J.~A., \& Fischer, D.~A.\ 2005, \apjs, 159, 141 

\bibitem[van Belle et al.(2008)]{GVB07} van Belle, G.~T., van Belle, G., Creech-Eakman, M.~J., et al.\ 2008, \apjs, 176, 276-292 

\bibitem[van Leeuwen(2008)]{HIP2} van Leeuwen, F., \ 2008, Hipparcos, the New Reduction, VizieR Online Data Catalog, 1311 

\bibitem[Young(1945)]{young45}
Young, R. K. 1945, Publ. of David Dunlap Obs., 1, 311

\end{thebibliography}
\end{document}